\documentclass[aps,12pt,preprint,prd]{revtex4}
\begin{document}

\draft
\title{$Qq\bar Q\bar q'$ States in Chiral SU(3) Quark Model}
\author{ZHANG Hai-Xia, ZHANG Min, ZHANG Zong-Ye}
\address{Institute of High Energy Physics, PO Box 918-4, Beijing 100049}

\begin{abstract}
We study the masses of $Qq\bar Q\bar q'$ states with
$J^{PC}=0^{++}$, $1^{++}$, $1^{+-}$ and $2^{++}$ in the chiral SU(3)
quark model, where $Q$ is the heavy quark ($c$ or $b$) and $q$
($q'$) is the light quark ($u$, $d$ or $s$). According to our
numerical results, it is improbable to make the interpretation of
$[cn\bar c\bar n]_{1^{++}}$ and $[cn\bar c\bar n]_{2^{++}}$ ($n=u,
d$) states as $X(3872)$ and $Y(3940)$, respectively. However, it is
interesting to find the tetraquarks in the $bq\bar b\bar q'$ system.


\end{abstract}
\maketitle

In the past three years, the discovery of few charmonium-like states
$X(3872)$,$^{[1]}$ $Y(3940)$$^{[2]}$  and $Y(4260)$$^{[3]}$  in
experiments has excited great interests of physicists. Three
different frameworks have been suggested to accommodate these states
with their unusual characteristics: (1) $D-D^*$ molecules;$^{[4]}$
(2) $\bar ccg$ hybrids,$^{[5]}$  and (3) diquark-antidiquark or
four-quark (4q) states for short.$^{[6]}$  Maiani {\it et
al.}$^{[6]}$ supposed that $X(3872)$ is a bound diquark-antidiquark
state with $J^{PC}=1^{++}$. With the spin-spin interactions and
$X(3872)$ mass as input, they predicted the existence of $2^{++}$
state at 3952 MeV that could be identified with $Y(3940)$. Using the
relativistic quark model, Ebert {\it et al.}$^{[6]}$ indicated that
$X(3872)$ could be the tetraquark state with hidden charm and the
masses of ground state tetraquarks with hidden bottom are below the
open bottom threshold. Assuming that $X(3872)$ is a $qc\bar q\bar c$
tetraquark and using its mass as input, Cui {\it et al.}$^{[6]}$
concluded that $0^+$ states will also exist if $X(3872)$ is really a
$1^+$ tetraquark. In this Letter, we study the $Qq\bar Q\bar q'$
states in the chiral SU(3) quark (CSQ) model, where heavy quark
$Q=c, b$ and light quark $q, q'=u, d$ or $s$.

Although the $QQ\bar q\bar q'$ states were studied in the
constituent quark model many years ago, no corresponding particles
were found in experiments. The discovery of such charmonium-like
states inspirits us to study the 4q states with content $Qq\bar
Q\bar q'$, moreover there is a lack of dynamics model calculation.
In our pervious work, we concluded that $D_s(2317)$ and $D_s(2460)$
were not the pure $cn\bar n\bar s$ 4q state in the CSQ
model,$^{[7]}$ which is consistent with the general opinions about
them. Thus it is interesting to study the $cq\bar c\bar q'$ states
again by using the same model parameters. Some authors$^{[8]}$ have
reported that systems with a large mass difference among their
components are more easily bound. Therefore, it is necessary to
study the $bq\bar b\bar q'$ states in the 4q-state picture.

The CSQ model is based on the constituent quark model of the light
quark systems, in which the constituent mass appears because of the
vacuum spontaneous breaking, and at the same time the coupling
between Goldstone bosons and quarks is automatically introduced for
restoring the chiral symmetry and these boson exchanges are
essential to obtain a correct description of the $NN$ phenomenology
and the light baryon spectrum. However, for the heavy quark systems,
their constituent part is very small and the vacuum spontaneous
breaking effect is not important. The one-gluon-exchange (OGE)
interaction between quarks and the confinement potential are enough
to describe the main properties of the heavy quark systems. This
basic framework is consistent with the QCD inspire and it is the
usual treatment in many works. For heavy-light quark systems, since
the masses of $D$, $D_s$, $B$ and $B_s$ mesons are quite large,
which locate inside the chiral symmetry scale, at least as the first
step, it is unnecessary to consider the Goldstone boson exchanges
between light and heavy quarks. Therefore, in this work, for the
light quark pairs, the interactions include confinement potential,
OGE potential and Goldstone boson exchanges; and for the heavy-heavy
and heavy-light quark interactions, the last one is not considered.
Explicit expressions of the interacting potential derived from the
nonrelativistic reduction of the Lagrangian on the static
approximation and a more detailed discussion of the model can be
found in Refs.\,[7,9].

The interaction parameters include the OGE coupling constant $g_i$,
the confinement strengths $(a_{ij}, a_{ij}^0)$, and the chiral
coupling constant $g_{ch}$. The parameters for the light quark pairs
are taken from our previous works,$^{[7,9]}$ which gave a
satisfactory description for the light baryon spectrum and the
binding energy of the deuteron. The parameters for $cq$ or $cc$
quark pairs are taken from Ref.\,[7], which fitted the masses of
$D$, $D^*$, $D_s$, $D_s^*$, $\eta _c$, $J/\Psi$ and $h_1(1p)$.
Followed the same method, the model parameters for $bq$ and $bb$
quark pairs can be fixed by the masses of $B$, $B^*$, $B_s$,
$B_s^*$, $\eta _b$ and $\Upsilon(1s)$, which are listed in Table 1,
and the theoretical results for the masses of $Q\bar q$ and $Q\bar
Q$ mesons are shown in Table 2. It is seen that the theoretical
masses of mesons are reasonably consistent with their experimental
values.
\begin{table}
\caption{Model parameters for the $bq$ and $bb$ quark pairs.}
\begin{center}
\begin{tabular}{lp{15mm}p{15mm}p{15mm}}
\hline \hline  &$m_b$(MeV) & & $g_b$  \\
& 4717 &  & 0.52 \\\hline
(MeV/fm)& $a_{bu}$ & $a_{bs}$ & $a_{bb}$ \\
& 275 & 275 & 275 \\
(MeV) & $a_{bu}^0$ & $a_{bs}^0$ & $a_{bb}^0$ \\
&-141.1 & -112 & -39.5\\\hline
\end{tabular}
\end{center}
\end{table}
\begin{table}
\caption{Masses (MeV) of $Q\bar q$ and $Q\bar Q$ mesons.
Experimental data are taken from PDG.}
\begin{center}
\begin{tabular}{p{15mm}p{12mm}p{12mm}p{12mm}p{12mm}l}
\hline \hline Mesons & $D$ & $D^{*}$ & $D_s$ & $D_s^{*}$ & \\
\hline
Exp.  & 1867.7 & 2008.9 & 1968.5 & 2112.4 &\\
Theor. & 1888 & 2009 & 1969 & 2130 & \\
\hline
\hline Mesons & $B$ & $B^{*}$ & $B_s$ & $B_s^{*}$ & \\
\hline
Exp.  & 5279.2 & 5325 & 5369.6 & 5416.6 & \\
Theor. & 5288 & 5320 & 5371 & 5412 & \\
\hline\hline Mesons & $\eta _c$ & $J/\Psi$ & $h_c(1p)$ & $\eta _b$ & $\Upsilon(1s)$ \\
\hline
Exp. & 2979.6 & 3096.9 & 3526.2 & 9300 & 9460.3\\
Theor. & 2990 & 3098 & 3568 & 9404 & 9460 \\\hline
\end{tabular}
\end{center}
\end{table}

For the wave function of a 4q state, let us describe it in the form
\[
\Psi (4q)=\psi _{4q}(0s^4)\left[ (q_1q_2)_{S_1,C_1}^{I_1}(\bar
q_3\bar q_4)_{S_2,C_2}^{I_2}\right] _{S,1^c}^I\;,
\]
where $\psi _{4q}(0s^4)$ is the orbital part and $\left[
(q_1q_2)_{S_1,C_1}^{I_1}(\bar q_3\bar q_4)_{S_2,C_2}^{I_2}\right]
_{S,C}^I$ is the flavor-spin-color part. As the first step, let us
take the spacial wave function of such four quarks in $S$ wave
state. Since there is no need to antisymmetrize the wave function
under the interchange of the coordinates of a heavy and a light
quark, the possible configurations of $Qq\bar Q\bar q'$ states with
$(J^P;I)=(1^{+};0)$ read
\[
\Psi _A=\psi _{Qq\bar Q\bar q'}(0s^4)\left[ (Qq)_{1,\bar 3^c}^{\frac
12}(\bar Q\bar q')_{0,3^c}^{\frac 12}\right] _{1,1^c}^0\,,
\]
\[
\Psi _B=\psi _{Qq\bar Q\bar q'}(0s^4)\left[ (Qq)_{1,6^c}^{\frac
12}(\bar Q\bar q')_{0,\bar 6^c}^{\frac 12}\right] _{1,1^c}^0\,,
\]
\[
\Psi _C =\psi _{Qq\bar Q\bar q'}(0s^4)\left[ (Qq)_{0,\bar
3^c}^{\frac 12}(\bar Q\bar q')_{1,3^c}^{\frac 12}\right] _{1,1^c}^0
,
\]
\[
\Psi _D =\psi _{Qq\bar Q\bar q'}(0s^4)\left[ (Qq)_{0,6^c}^{\frac
12}(\bar Q\bar q')_{1,\bar 6^c}^{\frac 12}\right] _{1,1^c}^0\,,
\]
\[
\Psi _E =\psi _{Qq\bar Q\bar q'}(0s^4)\left[ (Qq)_{1,\bar
3^c}^{\frac 12}(\bar Q\bar q')_{1,3^c}^{\frac 12}\right]
_{1,1^c}^0\,,
\]
\[
\Psi _F =\psi _{Qq\bar Q\bar q'}(0s^4)\left[ (Qq)_{1,6^c}^{\frac
12}(\bar Q\bar q')_{1,\bar 6^c}^{\frac 12}\right] _{1,1^c}^0\,.
\]
Note that both the configurations $[(Qq)_{\bar 3^c}(\bar Q\bar
q')_{3^c}]_{1^c}$ and $[(Qq)_{6^c}(\bar Q\bar q')_{\bar 6^c}]_{1^c}$
are taken into account in wave functions. Under charge conjugation,
$\Psi_A$ ($\Psi_B$) and $\Psi_C$ ($\Psi_D$) interchange while
$\Psi_E$ ($\Psi_F$) is odd. Thus, $J^P=1^+$ complex contains two
$C$-even and four $C$-odd states:
\begin{equation}
\begin{array}{l}
|J^{PC},I\rangle =|1^{++},0\rangle _1=\frac 1{\sqrt{2}}[\Psi_C+\Psi_A],\\
|1^{++},0\rangle _2=\frac 1{\sqrt{2}}[\Psi_D+\Psi_B],
\end{array}
%   (11)
\end{equation}
\begin{equation}
\begin{array}{l}
|1^{+-},0\rangle _1=\frac 1{\sqrt{2}}[\Psi_C-\Psi_A],\\
|1^{+-},0\rangle _2=\frac 1{\sqrt{2}}[\Psi_D-\Psi_B],\\
|1^{+-},0\rangle _3=\Psi_E,\\
|1^{+-},0\rangle _4=\Psi_F,
\end{array}
%   (11)
\end{equation}
where $|1^{++},0\rangle _1$ in Eq.\,(1) and $|1^{+-},0\rangle _1$
and $|1^{+-},0\rangle _3$ in Eq.\,(2) are the so-called
diquark-antidiquark states in Ref.\,[6]. In order to analyse the $C$
values of Eqs.\,(1) and (2), we recouple the color basis from
$\{|\bar 3_{Qq},3_{\bar Q\bar q'}\rangle , |6_{Qq},\bar 6_{\bar
Q\bar q'}\rangle \}$ to $\{|1_{Q\bar Q},1_{q\bar q'}\rangle ,
|8_{Q\bar q'},8_{q\bar Q}\rangle \}$. Take $|1^{++},0\rangle _1$ for
example,
\begin{eqnarray}
|1^{++},0\rangle _1 &=&\frac 1{\sqrt{3}}\psi _{Q\bar Qq\bar
q'}(0s^4)[(Q\bar Q)_{1,1^c}^0(q\bar
q')_{1,1^c}^0]_{1,1^c}^0 \nonumber\\
&&-\sqrt{\frac 23}\psi _{Q\bar Qq\bar q'}(0s^4)[(Q\bar
Q)_{1,8^c}^0(q\bar q')_{1,8^c}^0]_{1,1^c}^0\\
or&=&\frac 1{\sqrt{6}}\psi _{Q\bar q'q\bar Q}(0s^4)[(Q\bar
q')_{0,1^c}^{\frac
12}(q\bar Q)_{1,1^c}^{\frac 12}]_{1,1^c}^0 \nonumber\\
&&-\frac 1{\sqrt{6}}\psi _{Q\bar q'q\bar Q}(0s^4)[(Q\bar
q')_{1,1^c}^{\frac 12}(q\bar Q)_{0,1^c}^{\frac 12}]_{1,1^c}^0 \nonumber\\
&&-\frac 1{\sqrt{3}}\psi _{Q\bar q'q\bar Q}(0s^4)[(Q\bar
q')_{0,8^c}^{\frac 12}(q\bar Q)_{1,8^c}^{\frac 12}]_{1,1^c}^0
\nonumber \\
&&+\frac 1{\sqrt{3}}\psi _{Q\bar q'q\bar Q}(0s^4)[(Q\bar
q')_{1,8^c}^{\frac 12}(q\bar Q)_{0,8^c}^{\frac 12}]_{1,1^c}^0.
\end{eqnarray}
From Eq.\,(3), the only one with $C=+$ is that both spins of $Q\bar
Q$ and $q\bar q'$ pairs equal to 1, namely, $S_{Q\bar Q}=S_{q\bar
q'}=1$. Similarly, the states with $C=-$, $S_{Q\bar Q}$ and
$S_{q\bar q'}$ should be 0 and 1, or 1 and 0, respectively. After a
similar deducing, we can confirm that Eq.\,(2) are $C$-odd states.
For $Qq\bar Q\bar q'$ states with $(J^{PC};I)=(2^{++};0)$, the
possible configurations read
\begin{equation}
|2^{++},0\rangle _1=\psi _{Qq\bar Q\bar q'}(0s^4)\left[ (Qq)_{1,\bar
3^c}^{\frac 12}(\bar Q\bar q')_{1,3^c}^{\frac 12}\right]
_{2,1^c}^0\,,
\end{equation}
\begin{equation}
|2^{++},0\rangle _2=\psi _{Qq\bar Q\bar q'}(0s^4)\left[
(Qq)_{1,6^c}^{\frac 12}(\bar Q\bar q')_{1,\bar 6^c}^{\frac
12}\right] _{2,1^c}^0\,.
\end{equation}
For the other quantum numbers, such as $J^{PC}=0^{++}$ or isospin
$I=1$, the wave functions can be written with the same rule. For
saving space, we do not show them here.

By using the variation method, the energies of these states can be
obtained. For the 4q states with the same ($J^{PC}, I$), the
configuration mixture should be considered.

For the recently observed $X(3872)$ and $X(3940)$ resonances the $S$
wave, $cn\bar c\bar n$ assignment has been suggested with
$J^P=1^{++}$ and $2^{++}$, respectively.$^{[10]}$ Thus, we focus our
calculation on $J^{PC}=0^{++}$, $1^{++}$ and $2^{++}$. At the same
time, the $bq\bar b\bar q'$ states with the same quantum numbers are
studied. The masses of $cq\bar c\bar q'$ and $bq\bar b\bar q'$
states are calculated and the numerical results are presented in
Table 3.
\begin{table}
\caption{Masses (MeV) of $cq\bar c\bar q'$ and $bq\bar b\bar q'$
states with different $(J^{PC}, I)$.}
\begin{center}
\begin{tabular}{lcccc}
\hline \hline 4q states& $cn\bar c\bar n'$& $cs\bar c\bar s$ & $bn\bar b\bar n'$ & $bs\bar b\bar s$  \\
\hline
$(0^{++};0)$& 3956 & 4177 & 10347 & 10608 \\
$(0^{++};1)$& 3862 & --- & 10260 & --- \\
$(1^{++};0)$& 4047 & 4444 & 10395 & 10799 \\
$(1^{++};1)$& 4123 & --- & 10464 & --- \\
$(1^{+-};0)$& 4003 & 4271 & 10366 & 10639 \\
$(1^{+-};1)$& 3926 & --- & 10285 & --- \\
$(2^{++};0)$& 4047 & 4444 & 10395 & 10799  \\
$(2^{++};1)$& 4123 & --- & 10464 & --- \\
\hline
\end{tabular}
\end{center}
\end{table}

In principle, the allowed decay modes depend on the relationship
between the tetraquark mass and the sum of the masses of the
possible decay products. In the case of $[cn\bar c\bar n]_{(1^{++},
0)}$ states, according to Eqs.\,(1), (3), and (4), the possible
decay products are $J/\Psi+\omega (n\bar n)$ and $D+D^*$. Similarly,
for the $[cn\bar c\bar n]_{(1^{++}; 1)}$ states, they are $J/\Psi +
\rho$ and $D+D^*$. From Table 3, the masses of $[cn\bar c\bar
n']_{1^{++}}$ states are 4047\,MeV and 4123\,MeV for isospin equal
to 0 and 1, respectively. They are all above the threshold of $DD^*$
(3897 MeV). Therefore, $[cn\bar c\bar n']_{1^{++}}$ is broad and not
easy to be detected experimentally, and this result is not
consistent with the experiments of $X(3872)$. Thus, we think that
the 4q-state picture cannot be suitable to $X(3872)$ in our
calculation. For $X(3940)$, the same conclusion can be obtained,
which is different from the conclusion of Maiani {\it et
al.}$^{[6]}$ and Ebert {\it et al.},$^{[6]}$ who indicated that
$Y(3940)$ is a $2^{++}$ tetraquark state. On the other hand, in our
model the lightest scalar $0^{++}$ $cn\bar c\bar n'$ states are
above the threshold of $DD$ (3776\,MeV) and thus are broad, which is
consistent with the conclusion of Ebert {\it et al.}$^{[6]}$, but
different from the result of Maiani {\it et al.}$^{[6]}$ and Cui
{\it et al.}$^{[6]}$ In short, the $cq\bar c\bar q'$ states with
$J^{PC}=0^{++}$, $1^{++}$ and $2^{++}$ may not be tetraquark states
in our present calculation.

How about the $bq\bar b\bar q'$ states? Do they have any chance to
exist as tetraquark states? For states $[bn\bar b\bar n']_{1^{++}}$,
from Table 3, the masses are 10395\,MeV and 10464\,MeV for isospin
$I=0$ and 1, respectively, which is below the threshold of $BB^*$
(10608\,MeV). Thus, $[bn\bar b\bar n]_{(1^{++},0)}$ can only decay
to $\Upsilon(1s) + \omega (n\bar n)$, and $[bn\bar b\bar
n]_{(1^{++},1)}$ can go to $\Upsilon(1s) + \rho $. Although $[bn\bar
b\bar n']_{1^{++}}$ is not a bound state here, its mass is below the
threshold of $BB^*$. Therefore, we think that it may be a 4q state
and can be predicted to be narrow. Such state is worth being found
in experiment. Additionally, our numerical results show that other
$bq\bar b\bar q'$ states we have studied may be 4q states, for
instance, $[bn\bar b\bar n]_{0^{++}}$ and $[bq\bar b\bar
q']_{1^{+-}}$. Although the masses of such states are high, about
10\,GeV, we also expect them to be found in the future experiments.

Form Table 3, we note that the masses of $[Qq\bar Q\bar
q']_{1^{++}}$ and $[Qq\bar Q\bar q']_{2^{++}}$ with the same isospin
are equal to each other. As an example, the energies of every
configuration state for $[cn\bar c\bar n]_{(1^{++}; 0)}$ and
$[cn\bar c\bar n]_{(2^{++}, 0)}$ are given in Table 4. If only the
$|1^{++},0\rangle _1$ and $|2^{++},0\rangle _1$ states are taken
into account, the masses of such two states are different. However,
the configuration mixture should be considered for the state with
the same quantum numbers, unless the contribution of some
configuration states are not important. From Table 4, we also note
that the energy of $|1^{++},0\rangle _1$ in Eq.\,(1) is slightly
higher than the energy of $|1^{++},0\rangle _2$. This result is
model dependent, and the possible main cause is that the color
electrostatic field (CEF) energy is considered in our calculation.
Additionally, whether the heavy-light diquark is formed or not in
the heavy-light 4q systems is also an open question now. After
considering the configuration mixture, the masses of
$|1^{++},0\rangle$ and $|2^{++},0\rangle$ are the same as shown in
Table 3. The reason is as follows: Since $|1^{++},0\rangle$ and
$|2^{++},0\rangle$ have the same $C$ number, both spins of $c\bar c$
and $q\bar q$ pairs in such states equal to 1. If $E_1>E_2$, the
lower eigenvalue of $|1^{++},0\rangle$ is $2E_2-E_1$, where $E_1$
and $E_2$ are the expected values of Hamiltonian on states
$|1^{++},0\rangle _1$ and $|1^{++},0\rangle _2$, respectively. The
corresponding eigenvector is $|1^{++},0\rangle =\sqrt{\frac 13}
|1^{++},0\rangle _1+\sqrt{\frac 23}|1^{++},0\rangle _2$. By
calculating the contribution of the interaction potentials, we note
that only the component $[(c\bar c)_{1,1^c}^0; (n\bar
n)_{1,1^c}^0]_{1,1^c}^0$ contributes. Following the same process, we
can obtain $|2^{++},0\rangle =\sqrt{\frac 13} |2^{++},0\rangle
_1+\sqrt{\frac 23}|2^{++},0\rangle _2$, and the interesting
component is $[(c\bar c)_{1, 1^c}^0;(n\bar n)_{1,1^c}^0]_{2,1^c}^0$.
Since a $c\bar c$ ($n\bar n$) pair in such two states have the same
quantum numbers, the masses of $|1^{++},0\rangle $ and
$|2^{++},0\rangle $ are equal to each other. This result means that
the molecule picture may be more suitable to describe $X(3872)$ and
$Y(3940)$. Obviously, if the CEF energy is deleted in our
calculation, the mass of such two states will be different.
Possibly, this is the place to review whether considering the
contribution of CEF or not.
\begin{table}
\caption{Energies (MeV) of every configuration state for $cn\bar
c\bar n$ states with $(J^{PC},I)=(1^{++},0)$ and $(2^{++},0)$.}
\begin{center}
\begin{tabular}{lcccc}\hline\hline
%& $[cn\bar c\bar n]_{(1^{++};0)}$ &  & $[cn\bar c\bar n]_{(2^{++};0)}$&   \\\hline
state & $|1^{++},0\rangle _1$& $|1^{++},0\rangle _2$ &
$|2^{++},0\rangle _1$ & $|2^{++},0\rangle _2$
\\\hline
Energy & 4167 & 4125 & 4230 & 4151 \\
\hline
\end{tabular}
\end{center}
\end{table}

In summary, we have studied the masses of $Qq\bar Q\bar q'$ states
in the CSQ model, and attempted to give a reasonable interpretation
of $X(3892)$ and $X(3940)$. Our numerical results show that they may
not be explained as the pure 4q state, while the $bn\bar b\bar n'$
states with $J^{PC}=0^{++}$, $1^{++}$ and $1^{+-}$ may be
tetraquarks and they are worth being detected in experiments. In our
present calculation, the orbital wave function of the four quarks
are in the $S$ wave state. This is the simplest condition, and other
conditions are also allowed. Since our 4q states are roughly of
hadronic size, the contribution of annihilation mechanics should be
taken into account. Moreover, the two quark-antiquark cluster
structure is worth using in the study of $X(3872)$ and $Y(3940)$.
These aspects will be studied in the next step to improve our
calculations.

This work is supported in part by the National Natural Science
Foundation of China: No. 10475087.

\end{document}